\documentclass{ifacconf}
\pdfoutput=1 
\usepackage[authoryear]{natbib}      
\usepackage{graphicx}          
\usepackage{amsmath}
\usepackage{amssymb}
\usepackage{mathrsfs}
\usepackage{savesym}
\savesymbol{AND}
\usepackage{algorithmic}
\usepackage{algorithm}
\DeclareMathOperator{\supp}{supp}
\DeclareMathOperator{\card}{Card}

\DeclareMathOperator*{\argmin}{arg\,min}
\DeclareMathOperator{\tr}{Tr}
\DeclareMathOperator{\diag}{diag}
\begin{document}
\begin{frontmatter}
\title{Input design for Bayesian identification of non-linear state-space models}
\author[First]{Aditya Tulsyan},
\author[First]{Swanand R. Khare},
\author[First]{Biao Huang},
\author[Second]{R. Bhushan Gopaluni},
\author[First]{J. Fraser Forbes}
\thanks[footnote]{This article has been published in: Tulsyan, A, S.R. Khare, B. Huang, R.B. Gopaluni and J.F. Forbes (2013). Bayesian identification of non-linear state-space models: Part I- Input design.  In: \emph{Proceedings of the 10th IFAC International Symposium on Dynamics and Control of Process Systems}. Mumbai, India.}
\thanks[footnote]{This work was supported by the Natural Sciences and Engineering Research Council (NSERC), Canada.}
\address[First]{Department of Chemical and Materials Engineering, University of Alberta, Edmonton AB T6G-2G6, Canada (e-mail: \{tulsyan; khare; biao.huang; fraser.forbes\}@ ualberta.ca)}
\address[Second]{Department of Chemical and Biological Engineering, University of
British Columbia, Vancouver BC V6T-1Z3, Canada (e-mail: gopaluni@chbe.ubc.ca)}
\begin{abstract}
We propose an algorithm for designing optimal inputs for on-line Bayesian identification of stochastic non-linear state-space models. The proposed method relies on minimization of the posterior Cram\'er Rao lower bound derived for the model parameters, with respect to the input sequence. To render the optimization problem computationally tractable, the inputs are parametrized as a multi-dimensional Markov chain in the input space. The proposed approach is illustrated through a simulation example.
\end{abstract}
\end{frontmatter}

\section{Introduction}
\label{sec:S1}
Over the last decade, great progress has been made within the statistics community in overcoming the computational issues, and making Bayesian identification tractable for a wide range of complicated models arising in demographic and population studies, image processing, and drug response modelling (\cite{G1995}). A detailed exposition of Bayesian identification methods can be found in \cite{Kantas2009}. This paper is directed towards the class of on-line methods for Bayesian identification of stochastic non-linear SSMs, the procedure for which is briefly introduced here first. Let $\{X_t\}_{t\in\mathbb{N}}$ and $\{Y_t\}_{t\in\mathbb{N}}$ be ${\mathcal{X}} (\subseteq \mathbb{R}^{n})$ and ${\mathcal{Y}} (\subseteq \mathbb{R}^{m})$ valued stochastic processes, and let $\{u_t\}_{t\in\mathbb{N}}$ be the sequence of inputs in $\mathbb{R}^{p}$, such that the state $\{X_t\}_{t\in\mathbb{N}}$ is an unobserved or unmeasured process, with initial density $p_{\theta}(x)$ and transition density $p_{\theta}(x'|x,u)$:
\begin{equation}
\label{eq:E1}
X_0\sim p_{\theta}(x_0);X_{t+1}|(x_t,u_{t})\sim p_\theta(x_{t+1}|x_t,u_{t})~~(t\in\mathbb{N}).
\end{equation}
$\{X_t\}_{t\in\mathbb{N}}$ is an unobserved process, but is observed through $\{Y_t\}_{t\in\mathbb{N}}$, such that $\{Y_t\}_{t\in\mathbb{N}}$ is conditionally independent given $\{X_t, u_t\}_{t\in\mathbb{N}}$, with marginal density $p_{\theta}(y|x,u)$:
\begin{equation}
\label{eq:E2}
Y_{t}|(x_t,u_t)\sim p_\theta(y_t|x_t, u_t)\quad(t\in\mathbb{N}).  
\end{equation}
$\theta$ in (\ref{eq:E1}) and (\ref{eq:E2}) is a vector of unknown  model parameters, such that ${\theta} \in {\Theta}$ is an open subset of $\mathbb{R}^{q}$. All the densities are with respect to suitable dominating measures, such as Lebesgue measure. Although (\ref{eq:E1}) and (\ref{eq:E2}) represent a wide class of non-linear time-series models, the model form and the assumptions considered in this paper are given below
\begin{equation}
\label{eq:E3}
{X}_{t+1}={f}_t({X}_{t},u_{t},\theta_t,V_t);\quad {Y}_t={g}_t({X}_{t}, u_t, \theta_t,W_t),
\end{equation}
where ${\{\theta_{t+1}=\theta_t\}_{t\in\mathbb{N}}=\theta}$ is a vector of static parameters.
\begin{assum}
\label{A1}
$\{V_t\}_{t\in\mathbb{N}}$ and $\{W_t\}_{t\in\mathbb{N}}$ are mutually independent sequences of independent random variables known a priori in their distribution classes (e.g., Gaussian) and parametrized by a known and finite number of moments.
\end{assum}
\begin{assum} 
\label{A2}
$\{f_t;g_t\}_{t\in\mathbb{N}}$ are such that in the open sets $\mathcal{X}$ and $\Theta$, $\{f_t;g_t\}_{t\in\mathbb{N}}$ is $\mathcal{C}^k({\mathcal{X}})$ and $\mathcal{C}^k(\Theta)$, respectively, and in $\mathbb{R}^p$, $\{f_t;g_t\}_{t\in\mathbb{N}}$ is $\mathcal{C}^{k-1}({\mathbb{R}^p})$, and in $\mathbb{R}^n$ and $\mathbb{R}^m$, $\{f_t\}_{t\in\mathbb{N}}$ is $\mathcal{C}^{k-1}({{\mathbb{R}^n}})$, and $\{g_t\}_{t\in\mathbb{N}}$ is $\mathcal{C}^{k-1}({{\mathbb{R}^m}})$, where $k\geq2$. 
\end{assum}
\begin{assum}
\label{A3}
For any random sample $(x_{t+1},x_t,u_{t},\theta_t,v_t)$ ${\in\mathcal{X}\times\mathcal{X}\times\mathbb{R}^p\times\Theta\times\mathbb{R}^n}$ and $(y_t,x_t,u_t,\theta_t,w_t)\in\mathcal{Y}\times\mathcal{X}\times\mathbb{R}^p\times\Theta\times\mathbb{R}^m$ satisfying (\ref{eq:E3}), $\nabla_{v_t}f^T_t(x_t,u_{t},\theta_t,v_t)$ and $\nabla_{w_t}g^T_t(x_t,u_t,\theta_t,w_t)$ have rank $n$ and $m$, respectively, such that using implicit function theorem, $p_\theta(x_{t+1}|x_t,u_t)=p(V_t=\tilde{f}_t(x_t,u_{t},\theta_t,x_{t+1}))$ and $p_{\theta}(y_{t}|x_t,u_t)=p(W_t=\tilde{g}_t(x_t,u_t,\theta_t,y_{t}))$ do not involve any Dirac delta functions. 
\end{assum}

For a generic sequence $\{r_t\}_{t\in\mathbb{N}}$, let $r_{i:j}\triangleq\{r_i,r_{i+1},\dots,r_j\}$. Let $\theta^\star \in {\Theta}\subseteq\mathbb{R}^q$ be the true, but unknown parameter vector generating a measurement sequence ${\{Y_{1:t}=y_{1:t}\}_{t\in\mathbb{N}}}$ given ${\{u_{1:t}\}_{t\in\mathbb{N}}}$, such that $X_{t+1}|(x_t,u_{t})\sim p_{\theta^\star}(x_{t+1}|x_t,u_{t})$ and $Y_{t}|(x_t,u_t)\sim p_{\theta^\star}(y_t|x_t,u_t)$. In Bayesian identification of  (\ref{eq:E3}), the problem of estimating the parameter vector ${\theta^\star\in {\Theta}\subseteq\mathbb{R}^q}$ in real-time, given a sequence of input-output data $\{u_{1:t},y_{1:t}\}_{t\in\mathbb{N}}$ is formulated as a joint state and parameter estimation problem. This is done by ascribing a prior density ${\theta_0\sim p(\theta_0)}$, such that ${\theta^\star\in\supp{p(\theta_0)}}$, and computing $\{p(z_t|u_{1:t},y_{1:t})\}_{t\in\mathbb{N}}$, where: ${Z_t\triangleq\{X_t;~\theta_t\}}$ is a ${\mathcal{Z} (\subseteq \mathbb{R}^{s=n+q})}$ valued extended Markov process with $Z_0\sim p_{\theta_0}(x_0)p(\theta_0)$ and $Z_t|(z_{t-1},u_{t-1})\sim p_{\theta_{t-1}}(x_t|x_{t-1},u_{t-1})\delta_{\theta_{t-1}}$ $(\theta_t)$. The inference on $\{\theta_t\}_{t\in\mathbb{N}}$ then relies on the marginal posterior $\{p(\theta_t|u_{1:t}, y_{1:t})\}_{t\in\mathbb{N}}$. Note that by a judicious choice of the input sequence $\{u_{1:t}\}_{t\in\mathbb{N}}$, $\{p(z_t|u_{1:t}, y_{1:t})\}_{t\in\mathbb{N}}$ can be `steered' in order to yield $\{p(\theta_t|u_{1:t}, y_{1:t})\}_{t\in\mathbb{N}}$, which gives more accurate inference on $\{\theta_t\}_{t\in\mathbb{N}}$. This is called the input design problem for Bayesian identification or simply, the Bayesian input design problem. A detailed review on this subject can be found in \cite{C1995}.

Bayesian input design for linear and non-linear regression models is an active area of research (see \cite{H2012}, \cite{K2006}, \cite{M1995} and references cited therein); however, its extension to SSMs has been limited. Recently, Bayesian input design procedure for non-linear SSMs, where $\{X_t\}_{t\in\mathbb{N}}$ is completely observed was developed by \cite{T2012a}.  Despite the success with regression models, to the best of authors' knowledge, no known Bayesian input design methods are available for identification of stochastic non-linear SSMs. This is due to the unobserved state process $\{X_t\}_{t\in\mathbb{N}}$, which makes the design problem difficult to solve.

This paper deals with the Bayesian input design for identification of stochastic SSMs given in (\ref{eq:E3}). The proposed method is based on minimization of the posterior Cram\'er-Rao lower bound (PCRLB), derived by \cite{T1998}. First, we use Monte-Carlo (MC) methods to obtain an approximation of the PCRLB, and then parametrize the inputs as a multi-dimensional Markov chain in $\mathbb{R}^p$, to render the optimization problem computationally tractable. Markov-chain parametrization not only allows to include amplitude constraints on the input, it can be easily implemented using a standard PID controller or any other regulator. 
The notation used here is given next.
  
\textit{Notation:} ${\mathbb{N}:=\{1,2,\dots\}}$; ${\mathbb{N}_0:=
\{0\}\cup\mathbb{N}}$; $\mathbb{R}^{\rm s\times s}$ is the set of real-valued ${s\times s}$ matrices of cardinality ${\card(\mathbb{R}^{\rm s\times s})}$; ${\mathcal{S}^s\subset \mathbb{R}^{\rm s\times s}}$ is the space of symmetric matrices; $\mathcal{S}^s_{+}$ is the cone of symmetric positive semi-definite matrices in $\mathcal{S}^s$; and $\mathcal{S}_{++}^s$ is its interior. The partial order on $\mathcal{S}^s$ induced by $\mathcal{S}_{+}^s$ and $\mathcal{S}_{++}^s$ are denoted by $\succcurlyeq$ and $\succ$, respectively. $\mathbb{F}^{\rm s\times s}\subset\mathbb{R}^{\rm s\times s}$ is the set of $s\times s$ stochastic matrix, where $\mathbb{F}:=[0,1]$ and the sum of each row adds up to $1$. For $A\in\mathbb{R}^{\rm s\times s}$, $\tr[A]$ denotes its trace. For vectors $x\in\mathbb{R}^p$, $y\in\mathbb{R}^p$, and $z\in\mathbb{R}^p$, $x\leq y\leq z$ denotes element-wise inequality, and $\diag(y)\in\mathcal{S}^p$  is a $p\times p$ diagonal matrix with elements of $y\in\mathbb{R}^p$ as its diagonal entries. Finally, $\Delta^{y}_{x}\triangleq\nabla_{x}\nabla_{y}^T$ is a Laplacian and $\nabla_{x}\triangleq\left[\frac{\partial{}}{\partial{x}}\right]$ is a gradient. 
\section{Problem formulation}
\label{sec:S2}
Bayesian input design for regression models is a well studied problem in statistics (\cite{C1995}); wherein, the problem is often formulated as follows
\begin{align}
\label{eq:E4}
{\psi(u^\star_{1:N})=\max_{u_{1:N}\in\mathbb{R}^{pN}}\sum_{t=1}^{N}\mathbb{E}_{p(\theta_t,y_{1:t}|u_{1:t})}[\psi(Y_{1:t},u_{1:t},\theta_t)]}
\end{align}
where $\{u^\star_{1:N}\}_{N\in\mathbb{N}}$ is an $N$-step ahead optimal input sequence, and $\psi(\cdot)$ is a utility function. 
When inference on $\{\theta_t\}_{t\in\mathbb{N}}$ is of interest,  \cite{L1956} suggested using the mean-square error (MSE) as a utility function, such that
\begin{equation}
\label{eq:E5}
{\psi(u^\star_{1:N})=\max_{u_{1:N}\in\mathbb{R}^{pN}}\sum_{t=1}^{N}-\Phi(P^\theta_{{t|t}}(u_{1:t}))},
\end{equation}
where ${P^\theta_{{t|t}}(u_{1:t})=\mathbb{E}_{p(\theta_t,y_{1:t}|u_{1:t})}[(\theta_t-{\theta}_{t|t})(\theta_t-{\theta}_{t|t})^T]}$ is the MSE associated  with the parameter estimate given by ${\theta}_{t|t}=$ $\mathbb{E}_{p(\theta_t|u_{1:t}, y_{1:t})}[\theta_t]$, and ${\Phi:\mathcal{S}^q_{++}\rightarrow\mathbb{R}}$ is a test function.
\begin{rem}
\label{R1}
For the model considered in (\ref{eq:E3}), the marginal posterior density $\{p(\theta_t|u_{1:t}, y_{1:t})\}_{t\in\mathbb{N}}$, or the expectation with respect to it, does not admit any analytical solution, and thus, (\ref{eq:E5}) cannot be computed in closed form. ~~~~~~~~~~\qed
\end{rem}
\begin{rem}
\label{R2}
Methods such as SMC and MCMC can be used to approximate $\{p(\theta_t|u_{1:t}, y_{1:t})\}_{t\in\mathbb{N}}$; however, it makes the computation in (\ref{eq:E5}) formidable (\cite{K2006}). Moreover, the input $\{u^\star_{1:N}\}_{N\in\mathbb{N}}$ is optimal only for the Bayesian estimator used to approximate $\{p(\theta_t|u_{1:t}, y_{1:t})\}_{t\in\mathbb{N}}.$~~~~~~\qed
\end{rem}
To address the issues in Remarks \ref{R1} and \ref{R2}, we propose to define a lower bound on the MSE first, and minimize the lower bound instead. 
The PCRLB, derived  by \cite{T1998} provides a lower bound on the MSE associated with the estimation of $\{Z_t\}_{t\in\mathbb{N}}$ from $\{p(z_t|u_{1:t}, y_{1:t})\}_{t\in\mathbb{N}}$, and is given in the next lemma.
\begin{lem}
\label{L1}
Let ${\{Y_{1:t}=y_{1:t}\}_{t\in\mathbb{N}}}$ be an output sequence generated from (\ref{eq:E3}) using $\{u_{1:t}\}_{t\in\mathbb{N}}$, then the MSE associated with the estimation of $\{Z_t\}_{t\in\mathbb{N}}$ from $\{p(z_t|u_{1:t}, y_{1:t})\}_{t\in\mathbb{N}}$ is bounded from below by the following matrix inequality
\begin{equation}
\label{eq:E6}
P^z_{{t|t}}\triangleq\mathbb{E}_{p(z_{t}, y_{1:t}|u_{1:t})}[(Z_t-{Z}_{t|t})(Z_t-{Z}_{t|t})^T]\succcurlyeq [J_t^z]^{-1},
\end{equation}
where: ${{Z}_{t|t}=\mathbb{E}_{p(z_t|u_{1:t},y_{1:t})}[Z_t]}$ is an estimate of ${\{Z_t\}_{t\in\mathbb{N}}}$; ${P^z_{t|t}\triangleq\left[
  \begin{array}{cc}
    P^{x}_{t|t} & P^{x\theta}_{t|t} \\
    (P^{x\theta}_{t|t})^T & P^{\theta}_{t|t} \\
  \end{array}
\right]\in\mathcal{S}^{s}_{++}}$, ${J^z_{t}\triangleq\left[
  \begin{array}{cc}
    J_{t}^{x} & J_{t}^{x\theta} \\
    (J_{t}^{x\theta})^T & J_{t}^{\theta} \\
  \end{array}
\right]\in\mathcal{S}^{s}_{++}}$, ${[J_t^z]^{-1}\triangleq\left[
  \begin{array}{cc}
    L_{t}^x & L_{t}^{x\theta} \\
    (L_{t}^{x\theta})^T & L_{t}^\theta \\
  \end{array}
\right]\in\mathcal{S}^{s}_{++}}$ are  the MSE, posterior information matrix (PIM), and PCRLB, respectively.
\end{lem}
\begin{pf}
See \cite{T1998} for proof. ~~~~~~~~~~~~~~~\qed
\end{pf}
\begin{lem}
\label{L2}
A recursive approach to compute $\{J_t^z\}_{t\in\mathbb{N}}$ for (\ref{eq:E3}) under Assumptions \ref{A1} through \ref{A3} is given as follows
\begin{subequations}
\label{eq:E7}
\begin{align}
J^{x}_{t+1}&=H_t^{33}-(H_t^{13})^T[J_t^{x}+H_t^{11}]^{-1}H_t^{13};\label{eq:E7a}\\
J^{x\theta}_{t+1}&=(H_t^{23})^T-(H_t^{13})^T[J_t^{x}+H_t^{11}]^{-1}(J_t^{x\theta}+H_t^{12});\label{eq:E7b}\\
J^{\theta}_{t+1}&=J_t^{\theta}+H_t^{22}-(J_t^{x\theta}+H_t^{12})^T[J_t^{x}+H_t^{11}]^{-1}
\nonumber\\
&\times(J_t^{x\theta}+H_t^{12}),\label{eq:E7d}
\end{align}
\end{subequations}
where:
\begin{subequations}
\label{eq:E8}
\begin{align}
H_t^{11}&=\mathbb{E}_{\tilde{p}_{t+1}}[-\Delta_{X_t}^{X_t}\log{p_t}];\label{eq:E8a}\\
H_t^{12}&=\mathbb{E}_{\tilde{p}_{t+1}}[-\Delta_{X_t}^{\theta_t}\log p_t];\label{eq:E8b}\\
H_t^{13}&=\mathbb{E}_{\tilde{p}_{t+1}}[-\Delta_{X_t}^{X_{t+1}}\log p_t];\label{eq:E8c}\\
H_t^{22}&=\mathbb{E}_{\tilde{p}_{t+1}}[-\Delta_{\theta_t}^{\theta_t}\log{p}_{t}];\label{eq:E8d}\\
H_t^{23}&=\mathbb{E}_{\tilde{p}_{t+1}}[-\Delta_{\theta_t}^{X_{t+1}}\log{p}_{t}];\label{eq:E8e}\\
H_t^{33}&=\mathbb{E}_{\tilde{p}_{t+1}}[-\Delta_{X_{t+1}}^{X_{t+1}}\log{p}_{t}];\label{eq:E8f}
\end{align}
\end{subequations}
${\tilde{p}_{t+1}=p(x_{0:t+1},\theta_t,y_{1:t+1}|u_{1:t+1})}$, and ${{p}_{t}=p(X_{t+1}|Z_t,u_t)}$ $p(Y_{t+1}|\theta_t,X_{t+1},u_{t+1})$; and $J_0=\mathbb{E}_{p(z_0)}[-\Delta_{Z_0}^{Z_0}\log{p}(Z_0)]$.
\end{lem}
\begin{pf}
See \cite{T1998} for proof.\qquad\qquad\qquad\qed
\end{pf}
\begin{cor}
\label{C1}
Let ${P^z_{{t|t}}\in\mathcal{S}^{s}_{++}}$, ${[J_t^z]^{-1}\in\mathcal{S}^{s}_{++}}$ be such that they satisfy (\ref{eq:E6}), then the MSE associated with the point estimation of $\{\theta_t\}_{t\in\mathbb{N}}$, computed from $\{p(\theta_t|u_{1:t},y_{1:t})\}_{t\in\mathbb{N}}$, is bounded from below by the following matrix inequality
\begin{equation}
\label{eq:E9}
P^\theta_{{t|t}}=\mathbb{E}_{p(\theta_{t}, y_{1:t}|u_{1:t})}[(\theta_t-{\theta}_{t|t})(\theta_t-{\theta}_{t|t})^T]\succcurlyeq L_t^\theta,
\end{equation}
where $L_t^\theta\in\mathcal{S}^{q}_{++}$ is the lower-right sub-matrix of $[J_t^z]^{-1}\in\mathcal{S}^{s}_{++}$ in (\ref{eq:E6}).
\end{cor}
\begin{pf}
The proof is based on the fact that the PCRLB inequality in (\ref{eq:E6}) guarantees that $P^z_{{t|t}}-[J_t^z]^{-1}\in\mathcal{S}^{s}_{+}$. ~~~\qed
\end{pf}
\begin{thm}
\label{T1}
Let $J_t^z\in\mathcal{S}^{s}_{++}$ be the PIM for model in (\ref{eq:E3}) and $L_t^\theta\in\mathcal{S}^{q}_{++}$ be the lower bound on the MSE associated with the estimation of $\{\theta_t\}_{t\in\mathbb{N}}$ in (\ref{eq:E3}), then given $J^z_t\in\mathcal{S}^{s}_{++}$, the lower bound $L_t^\theta\in\mathcal{S}^{q}_{++}$ at $t\in\mathbb{N}$ can be computed as
\begin{equation}
\label{eq:E10}
L_t^{\theta}=[J^\theta_t-(J_t^{x\theta})^T(J_t^{x})^{-1}J_t^{x\theta}]^{-1},
\end{equation}
where $J^\theta_t$, $J_t^{x\theta}$ and $J_t^{x}$ are the PIMs given in {Lemma \ref{L2}}.
\end{thm}
\begin{pf}
The proof is based on matrix inversion lemma.\qed
\end{pf}
Finally, the input design problem for Bayesian identification of $\{\theta_{t}\}_{t\in\mathbb{N}}$ in (\ref{eq:E3}) can  be formulated as follows
\begin{subequations}
\label{eq:E11}
\begin{align}
\psi(u^\star_{1:N})=&\min_{u_{1:N}\in\mathbb{R}^{pN}}\sum_{t=1}^{N}\Phi(L^\theta_{{t}}(u_{1:t}))\\
&~~~~\text{s.t.}~~~u_{min}\leq \{u_{i}\}_{t\in[1,N]}\leq u_{max},
\end{align}
\end{subequations}
where $L^\theta_{{t}}(u_{1:t})\triangleq L^\theta_{{t}}$; and $u_{max}\in\mathbb{R}^p$ and $u_{min}\in\mathbb{R}^p$ are the maximum and minimum magnitude of the input.
\begin{rem}
\label{R5}
The optimization problem in (\ref{eq:E11}) allows to impose magnitude constraints on the inputs. Although constraints on $(x_{0:N})\in\mathcal{X}^{N+1}$ and $(y_{1:N})\in\mathcal{Y}^N$ are not included, but if required, they can also be appended.~~~~~~~~\qed
\end{rem}
\begin{rem}
\label{R4}
Integral in (\ref{eq:E8}), with respect to $\tilde{p}_t$, makes (\ref{eq:E11}) independent of the random realizations from $\mathcal{X}^{t+1}$, $\Theta$, and $\mathcal{Y}^{t}$. The optimization in (\ref{eq:E11}) in fact only depends on: the process dynamics represented in (\ref{eq:E3}); noise densities $V_t\sim p(v_t)$ and $W_t\sim p(w_t)$; and the choice of  $Z_0\sim p(z_0)$ and ${u_{1:N}}\in\mathbb{R}^{pN}$. This makes (\ref{eq:E11})  independent of ${\theta^\star\in\Theta\subseteq\mathbb{R}^q}$ or the Bayesian estimator used for estimating ${\{\theta_t\}_{t\in\mathbb{N}}}$.
~~\qed 
\end{rem}
\begin{rem}
\label{R7}
The formulation in  (\ref{eq:E11}) yields a sequence $\{u^\star_{1:N}\}_{N\in\mathbb{N}}$, which is (a) optimal for all the Bayesian identification methods that approximate $\{p(\theta_t|{u^\star_{1:t},y_{1:t}})\}_{t\in\mathbb{N}}$; and (b) independent of $\theta^\star\in\mathbb{R}^q$  (see Remark \ref{R4}), such that the input $\{u^\star_{1:N}\}_{N\in\mathbb{N}}$ is optimal for all ${\theta^\star\in\supp{p(\theta_0)}}$.\qed
\end{rem}
There are two challenges that need to be addressed in order to make the optimization problem in (\ref{eq:E11}) tractable: (a)  computing the lower bound $\{L^\theta_{t}\}_{t\in\mathbb{N}}$; and (b) solving the high-dimensional optimization problem in $\mathbb{R}^{pN}$. 
Our approach to address the above challenges is discussed next.

\section{Computing the lower bound}
The first challenge is to compute the lower bound $L_t^\theta$ in (\ref{eq:E11}). It is well known that computing $L_t^\theta$ in closed form is  non-trivial for the model form considered in (\ref{eq:E3}) (see \cite{T1998},  \cite{NB2001}). This is because of the complex, high-dimensional integrals in (\ref{eq:E8a}) through (\ref{eq:E8f}), which do not admit any analytical solution.

MC sampling is a popular numerical method to solve integrals of the form ${F(u_{1:t})=\mathbb{E}_{p(x_{0:t}|u_{1:t})}[h(X_{0:t},u_{1:t})]}$, where ${h:\mathcal{X}^{t+1}\times\mathbb{R}^{pt}\rightarrow \mathbb{R}}$. Using $M$ i.i.d. trajectories ${\{X^{i}_{0:t}|u_{1:t}\}_{i=1}^{M}\sim p(x_{0:t}|u_{1:t})}$, the probability distribution ${p(x_{0:t}|u_{1:t})dx_{0:t}\triangleq p(dx_{0:t}|u_{1:t})}$, can be approximated as
\begin{equation}
\label{eq:E12}
\tilde{p}(dx_{0:t}|u_{1:t})=\frac{1}{M}\sum_{i=1}^M \delta_{X^i_{0:t}|u_{1:t}}(dx_{0:t}),
\end{equation} 
where $\tilde{p}(dx)$ is a MC estimate of ${p}(dx)$ and $\delta_{x_0}(dx)$ is the Dirac delta mass at $x_0$. Finally, substituting (\ref{eq:E12}) into $F(u_{1:t})$, we get $\tilde{F}(u_{1:t})\triangleq\tilde{F}(\{X^i_{0:t}|u_{1:t}\}_{i=1}^M)=\int h(x_{0:t},u_{1:t})\tilde{p}(dx_{0:t}|u_{1:t})=\frac{1}{M}\sum_{i=1}^M h(X^{i}_{0:t},u_{1:t})$,
where $\tilde{F}(u_{1:t})$ is an $M$-sample MC estimate of ${F}(u_{1:t})$.
\begin{rem}
\label{R11}
Using MC methods, the multi-dimensional integrals in (\ref{eq:E8a}) through (\ref{eq:E8f}), with respect to the density $\tilde{p}_t(\cdot)$ can be approximated by simulating $M$ i.i.d. sample paths $\{X^i_{0:t},\theta^i_t,Y^i_{1:t}|u_{1:t}\}_{i=1}^M\sim \tilde{p}_t(\cdot)$ using (\ref{eq:E3}), starting at $M$ i.i.d. initial positions drawn from $\{Z^i_0\}_{i=1}^M\sim p(z_0)$.  
\end{rem}
\begin{exmp}
\label{Ex1}
Consider the following stochastic SSM with additive Gaussian state and measurement noise
\begin{subequations}
\label{eq:E13}
\begin{align}
X_{t+1}&=f_t(X_t,\theta_t,u_t)+V_t,\label{eq:E13a}\\
Y_{t}&=g_t(X_t,\theta_t,u_t)+W_t,\label{eq:E13b}
\end{align}
\end{subequations}
where $\{V_t\}_{t\in\mathbb{N}}$ and $\{W_t\}_{t\in\mathbb{N}}$ are mutually independent sequences of independent zero mean Gaussian random variables, such that $V_t\sim\mathcal{N}(0,Q_t)$ and $W_t\sim\mathcal{N}(0,R_t)$, where $Q_t<\infty$ and $R_t<\infty$ for all $t\in\mathbb{N}$.
\end{exmp}
Note that for the model form considered in Example \ref{Ex1}, using the Markov property of the states and conditional independence of the measurements, the dimension of the integrals in (\ref{eq:E8a}) through (\ref{eq:E8f}) can be reduced, as given next.
\begin{thm}
\label{T2}
For a stochastic non-linear SSM in Example \ref{Ex1}, the matrices in (\ref{eq:E8a}) through (\ref{eq:E8f}) can be written as
\begin{subequations}
\label{eq:E14}
\begin{align}
H_t^{11}&=\mathbb{E}_{p(x_{t},\theta_t|u_{1:t+1})}[\nabla_{X_t}f_t^T(X_t,\theta_t,u_t)]Q_t^{-1}\nonumber\\
&\times[\nabla_{X_t}f_t^T(X_t,\theta_t,u_t)]^T;\label{eq:E14a}\\
H_t^{12}&=\mathbb{E}_{p(x_{t},\theta_t|u_{1:t+1})}[\nabla_{X_t}f_t^T(X_t,\theta_t,u_t)]Q_t^{-1}\nonumber\\
&\times[\nabla_{\theta_t}f_t^T(X_t,\theta_t,u_t)]^T;\label{eq:E14b}\\
H_t^{13}&=-\mathbb{E}_{p(x_{t},\theta_t|u_{1:t+1})}[\nabla_{X_t}f_t^T(X_t,\theta_t,u_t)]Q_t^{-1};\label{eq:E14c}\\
H_t^{22}&=\mathbb{E}_{p(x_{t},\theta_t|u_{1:t+1})}[\nabla_{\theta_t}f_t^T(X_t,\theta_t,u_t)]Q_t^{-1}\nonumber\\
&\times[\nabla_{\theta_t}f_t^T(X_t,\theta_t,u_t)]^T\nonumber\\
&+\mathbb{E}_{p(x_{t+1},\theta_t|u_{1:t+1})}[\nabla_{\theta_t}g_t^T(X_{t+1},\theta_t,u_{t+1})]R_{t+1}^{-1}\nonumber\\
&\times[\nabla_{\theta_t}g_t^T(X_{t+1},\theta_t,u_{t+1})]^T\label{eq:E14d}\\
H_t^{23}&=-\mathbb{E}_{p(x_{t},\theta_t|u_{1:t+1})}[\nabla_{\theta_t}f_t^T(X_t,\theta_t,u_t)]Q_t^{-1}\nonumber\\
&+\mathbb{E}_{p(x_{t+1},\theta_t|u_{1:t+1})}[\nabla_{\theta_t}g_t^T(X_{t+1},\theta_t,u_{t+1})]R_{t+1}^{-1}\nonumber\\
&\times[\nabla_{X_{t+1}}g_t^T(X_{t+1},\theta_t,u_{t+1})]^T\label{eq:E14e}\\
H_t^{33}&=Q_t^{-1}+\mathbb{E}_{p(x_{t+1},\theta_t|u_{1:t+1})}[\nabla_{X_{t+1}}g_t^T(X_{t+1},\theta_t,u_{t+1})]\nonumber\\
&\times R_{t+1}^{-1}[\nabla_{X_{t+1}}g_t^T(X_{t+1},\theta_t,u_{t+1})]^T\label{eq:E14f}
\end{align}
\end{subequations}
\end{thm}
\begin{pf}
(\ref{eq:E14a}): First note that $H_t^{11}=\mathbb{E}_{\tilde{p}_{t+1}}[-\Delta_{X_t}^{X_t}\log{p_t}]=\mathbb{E}_{\tilde{p}_{t+1}}[\nabla_{X_t}\log{p_t}][\nabla_{X_t}\log{p_t}]^T$ (see \cite{T1998}). On simplifying, we have $H_t^{11}=\mathbb{E}_{p(x_{t+1}|x_t,\theta_t,u_t)P(x_t,\theta_t|u_{1:t+1})}$ $[\nabla_{X_t}\log p(X_{t+1}|X_t,\theta_t,u_t)][\nabla_{X_t}\log p(X_{t+1}|X_t,\theta_t,u_t)]^T$.\newline This is due to $\nabla_{X_t}\log p(Y_{t+1}|X_{t+1},\theta_t,u_{t+1})=0$.  For Example \ref{Ex1}, $\nabla_{X_t}\log p(X_{t+1}|X_t,\theta_t,u_t)$ $=[\nabla_{X_t}f_t^T(X_t,\theta_t,u_t)]$ $Q_t^{-1}[X_{t+1}-f_t(X_t,\theta_t,u_t)]^T$. Substituting it into $H_t^{11}$, and using  $E_{p(x_{t+1}|x_t,\theta_t,u_t)}[X_{t+1}-f_t(X_t,\theta_t,u_t)][X_{t+1}-f_t(X_t,\theta_t,u_t)]^T$ $=Q_{t}$, we have (\ref{eq:E14a}). Note that the expression in (\ref{eq:E14b}) through (\ref{eq:E14f}) can be similarly derived.~~~~~~~\qed
\end{pf}
Theorem \ref{T2} reduces the dimension of the integral in (\ref{eq:E8}) for Example \ref{Ex1} from $(t+1)(n+m)+s$ to $s$. Using MC sampling, (\ref{eq:E14a}), for instance, can be computed as
$\tilde{H}_t^{11}=\frac{1}{M}\sum_{i=1}^M[\nabla_{X_t}f^T(X^i_t,\theta^i_t,u_t)]Q_t^{-1}[\nabla_{X_t}f^T(X^i_t,\theta^i_t,u_t)]^T$. Here $\{X_t^{i},\theta_t^i|u_{1:t+1}\}_{i=1}^M\sim p(x_t,\theta_t|u_{1:t+1})$ and $\tilde{H}_t^{11}$ is an $M$-sample MC estimate of ${H}_t^{11}$. Note that the MC estimates of (\ref{eq:E14b}) through (\ref{eq:E14f}) can be similarly computed. 
In general, substituting the MC estimates of (\ref{eq:E8a}) through (\ref{eq:E8f}) first into Lemma \ref{L2}, and then into Theorem \ref{T1}, yields  
\begin{equation}
\label{eq:E25}
\tilde{L}_t^{\theta}=[\tilde{J}^\theta_t-(\tilde{J}_t^{x\theta})^T(\tilde{J}_t^{x})^{-1}\tilde{J}_t^{x\theta}]^{-1},
\end{equation}
where $\tilde{L}^\theta_t$ is an estimate of ${L}^\theta_t$, and $\tilde{J}^\theta_t$, $\tilde{J}_t^{x\theta}$ and $\tilde{J}_t^{x}$ are the estimates of the PIMs in Lemma \ref{L2}. 
Finally, substituting (\ref{eq:E25}) into (\ref{eq:E11}) gives the following optimization problem
\begin{subequations}
\label{eq:E17a}
\begin{align}
\tilde{\psi}(u^\star_{1:N})=&\min_{u_{1:N}\in\mathbb{R}^{pN}}\sum_{t=1}^{N}\Phi(\tilde{L}^\theta_{{t}}(u_{1:t}))\\
&~~~~\text{s.t.}~~~u_{min}\leq \{u_{i}\}_{t\in[1,N]}\leq u_{max}.
\end{align}
\end{subequations}
\begin{thm}
\label{T4}
Let ${\psi}(u^\star_{1:N})$ and $\tilde{\psi}(u^\star_{1:N})$ be the optimal utility functions, computed by solving the optimization problem in (\ref{eq:E11}) and (\ref{eq:E17a}), respectively, then we have 
\begin{equation}
\label{eq:E17b}
\tilde{\psi}(u^\star_{1:N})\xrightarrow[M\rightarrow +\infty] {a.s.}{\psi}(u^\star_{1:N}),
\end{equation}
where $\xrightarrow{a.s.}$ denotes almost sure convergence.
\end{thm} 
\begin{pf}
Since (\ref{eq:E25}) is based on perfect MC sampling, using the strong law of large numbers, we have $\tilde{L}_t^{\theta}\xrightarrow[]{a.s.}{L}_t^{\theta}$ as $M\rightarrow +\infty$. Equation (\ref{eq:E17b}) follows from this result, which completes the proof.~~~~~~~~~~~~~~~~~~~~~~~~~~~~~~~~~~~~~~~~~~~~~~~~\qed
\end{pf}
A natural approach to solve (\ref{eq:E17a}) is to treat ${\{u_{1:N}\}_{N\in\mathbb{N}}}$ as a vector of continuous variables in ${\mathbb{R}^{pN}}$; however, this will render (\ref{eq:E17a}) computationally inefficient for large ${N\in\mathbb{N}}$. A relaxation method to make (\ref{eq:E17a}) tractable is given next. 
\section{Input parametrization}
\label{sec:S4}
To overcome the complications due to continuous valued input ${\{u_t\}_{t\in\mathbb{N}}\in\mathbb{R}^p}$, we discretize the input space from $\mathbb{R}^p$ to ${\mathcal{U}\subseteq\mathbb{R}^p}$, such that ${\card (\mathcal{U})=r}$, where ${r=b^p}$, and ${b\in\mathbb{N}}$ is the number of discrete values for each input in $\mathbb{R}$. If we denote ${\mathcal{U}=\{s_1,\dots,s_{r}\}}$, then ${u_{min}\leq s_i\leq u_{max}}$, for all ${1\leq i\leq r}$, such that (\ref{eq:E17a}) can be written as follows
\begin{equation}
\label{eq:E17}
\tilde{\psi}(u^\star_{1:N})=\min_{u_{1:N}\in\mathcal{U}^{N}}\sum_{t=1}^{N}\Phi(\tilde{L}^\theta_{{t}}(u_{1:t})).
\end{equation}
Note that although the input $\{u_{1:N}\}_{N\in\mathbb{N}}$ in (\ref{eq:E17}) is defined on a discrete input space $\mathcal{U}^{N}$ of $\card(\mathcal{U}^{N})=r^N$, (\ref{eq:E17}) is still intractable for large ${N\in\mathbb{N}}$. To address this issue, a multi-dimensional Markov chain input parametrization, first proposed by \cite{B2009}, is used here.
\begin{defn}
\label{D1}
For ${k\in\mathbb{N}_0}$ and ${\mathbb{S}:=\{k+1,k+2,\cdots\}}$, let  ${\{{U}_t\}_{t\in\mathbb{S}}=\{u_{t-k:t}\}_{k\in\mathbb{N}_0}}$ be a $\mathcal{U}^{k+1}$ valued first-order finite Markov chain, where ${\card(\mathcal{U}^{k+1})=r^{k+1}}$, such that the sample values of ${\{{U}_t\}_{k\in\mathbb{N}_0,t\in\mathbb{S}\setminus\{k+1\}}\in\mathcal{U}^{k+1}}$,  depend on the past only through the sample values of ${\{{U}_{t-1}\}_{t-1\in\mathbb{S}}\in}$ ${\mathcal{U}^{k+1}}$, such that for all ${\{{U}_t\}_{k\in\mathbb{N}_0,t\in\mathbb{S}\setminus\{k+1\}}\in\mathcal{U}^{k+1}}$ and ${\{{U}_{k+1:t-1}\}_{k\in\mathbb{N}_0, t-1\in\mathbb{S}}\in\mathcal{U}^{t-1}}$, we have the following
\begin{align}
\label{eq:E18}
\Pr({U}_{t}&=\{u_{t-k:t}\}|{{U}_{k:t-1}=\{u_{1:t-1}}\})=\nonumber\\
&P_{\Pi}({{U}_{t}=\{u_{t-k:t}\}}|{{U}_{t-1}=\{u_{t-k-1:t-1}\}}),
\end{align}
where $\Pr(\cdot)$ is a probability measure and $P_{\Pi}\in\mathbb{F}^{\rm r^{k+1}\times r^{k+1}}$ is a $r^{k+1}\times r^{k+1}$ probability transition matrix.~~~~~~~~~~~~~~~~~~~~~\qed 
\end{defn}
In Definition \ref{D1}, $P_{\Pi}({U_{t}=s_2})|{U_{t-1}=s_1})$, where $\{s_1,s_2\}$ $\in\mathcal{U}^{k+1}$ represents the probability that the Markov chain transits from ${\{{U}_{t-1}\}_{k\in\mathbb{N}_0,t-1\in\mathbb{S}}=s_1}$ to the input state  ${\{{U}_{t}\}_{k\in\mathbb{N}_0,t\in\mathbb{S}\setminus\{k+1\}}=s_2}$. Consider the following example.
\begin{exmp}
\label{E2}
For ${p=1}$, ${k=0}$, and ${b\in\mathbb{N}}$, we have ${r=b}$ and ${\mathbb{S}=\mathbb{N}}$, such that ${\{U_t\}_{t\in\mathbb{S}}=\{u_{t}\}}$ is a Markov chain on the input space ${\mathcal{U}=\{s_1,s_2,\dots,s_b\}}$ of ${\card(\mathcal{U})=b}$, then the probability matrix $P_{\Pi}\in\mathbb{F}^{\rm b\times b}$ can be represented as
\begin{equation}
\label{eq:E19}
P_{\Pi}=\left[
  \begin{array}{cccc}
    p_{s_1, s_1} & p_{s_1, s_2}&\cdots&p_{s_1, s_b} \\
    p_{s_2, s_1} & p_{s_2, s_2}&\cdots&p_{s_2, s_b} \\
    \vdots&\vdots&&\vdots\\
    p_{s_b, s_1} & p_{s_b, s_2}&\cdots&p_{s_b, s_b} \\
  \end{array}
\right],\nonumber
\end{equation}
where ${p_{s_i, s_j}}\triangleq P_{\Pi}({U_{t}=s_j})|{U_{t-1}=s_i})~\forall 1\leq i,j\leq b$.
\end{exmp}
\begin{exmp}
For ${p=1}$, ${k=1}$, and ${b\in\mathbb{N}}$, we have ${r=b}$ and ${\mathbb{S}=\mathbb{N}\setminus\{1\}}$, such that $\{U_t\}_{t\in\mathbb{S}}=\{u_{t-1:t}\}$ is a Markov chain on $\mathcal{U}^2=\{\{s_1,s_1\},\{s_1,s_2\},\dots,\{s_2,s_1\},\dots,\{s_b,s_b\}\}$ of $\card(\mathcal{U}^2)=b^2$ then $P_{\Pi}\in\mathbb{F}^{\rm b^2\times b^2}$ can be represented as
\begin{align*}
P_{\Pi}=\left[
  \begin{array}{cccc}
    p_{\{s_1,s_1\},\{s_1,s_1\}} & p_{\{s_1,s_1\},\{s_1,s_2\}}&\cdots&p_{\{s_1,s_1\},\{s_b,s_b\}} \\
    p_{\{s_1,s_2\},\{s_1,s_1\}} & p_{\{s_1,s_2\},\{s_1,s_2\}}&\cdots&p_{\{s_1,s_2\},\{s_b,s_b\}} \\
    \vdots&\vdots&&\vdots\\
    p_{\{s_1,s_g\},\{s_1,s_1\}} & p_{\{s_1,s_g\},\{s_1,s_2\}}&\cdots&p_{\{s_1,s_g\},\{s_g,s_g\}} \\
    p_{\{s_2,s_1\},\{s_1,s_1\}} & p_{\{s_2,s_1\},\{s_1,s_2\}}&\cdots&p_{\{s_2,s_1\},\{s_g,s_g\}} \\
    \vdots&\vdots&&\vdots\\
     \vdots&\vdots&&\vdots\\
    p_{\{s_g,s_g\},\{s_1,s_1\}} & p_{\{s_g,s_g\},\{s_1,s_2\}}&\cdots&p_{\{s_g,s_g\},\{s_g,s_g\}} \\
  \end{array}
\right].
\end{align*}  
where ${p_{\{s_i, s_j\},\{s_l, s_m\}}\triangleq P_{\Pi}({U_{t}=\{s_i, s_j\}})|{U_{t-1}=\{s_l, s_m\}})}$ $\forall 1\leq i,j,l,m\leq b$.
\end{exmp}
\begin{assum}
\label{A4}
The Markov chain $\{{U}_t\}_{t\in\mathbb{S}}=\{u_{t-k:t}\}_{k\in\mathbb{N}_0}$ considered in Definition \ref{D1} is time-homogeneous.
\end{assum}
\begin{assum}
\label{A5}
The Markov chain $\{{U}_t\}_{t\in\mathbb{S}}=\{u_{t-k:t}\}_{k\in\mathbb{N}_0}$ in Definition \ref{D1} has a prior probability distribution ${U_{k+1}\sim P_{\Gamma}(\{u_{1:k+1}\})}$, where ${P_{{\Gamma}}}$ is a $1\times r^{k+1}$ vector.
\end{assum}
\begin{thm}
\label{T3}
For ${k\in\mathbb{N}_0}$ and ${\mathbb{S}:=\{k+1,k+2,\cdots\}}$, let ${\{{U}_t\}_{t\in\mathbb{S}}=\{u_{t-k:t}\}_{k\in\mathbb{N}_0}}$ be a Markov chain defined in Definition \ref{D1}, and satisfying Assumptions \ref{A4} and \ref{A5}, such that ${U_{t}|(\{u_{t-k-1:t-1}\})\sim P_{\Pi}({\{u_{1:k+1}\}|\{u_{t-k-1:t-1}\}})}$ for all ${t\in\mathbb{S}\setminus\{k+1\}}$ and ${U_{k+1}\sim P_{\Gamma}(\{u_{1:k+1}\})}$ then ${\{U_{k+1:N}\}_{N\in\mathbb{N}}}\sim P_{\Gamma,\Pi}^{k+1:N}$ has a probability distribution
\begin{align}
\label{eq:E20}
{P_{\Gamma}(\{u_{1:k+1}\})}\prod_{t=k+2}^N P_{\Pi}({\{u_{t-k:t}\}|\{u_{t-k-1:t-1}\}}).
\end{align}
\end{thm}
\begin{pf}
Using probability chain rule, the joint probability distribution of ${U_{k+1:N}\sim P_{\Gamma,\Pi}^{k+1:N}}$ can be written as
\begin{subequations}
\begin{align}
&P_{\Gamma,\Pi}^{k+1:N}=\Pr(\{u_{1:k+1}\},\{u_{2:k+2}\},\dots,\{u_{N-k:N}\})\nonumber\\
&=\Pr(\{u_{N-k:N}\}|\{u_{1:k+1}\},\{u_{2:k+2}\},\dots,\{u_{N-k-1:N-1}\})\nonumber\\
&\times \Pr(\{u_{1:k+1}\},\{u_{2:k+2}\},\dots,\{u_{N-k-1:N-1}\}),\label{eq:E21a}\\
&=P_{\Pi_{k,r}}(\{u_{N-k:N}\}|\{u_{N-k-1:N-1}\})\nonumber\\
&\times \Pr(\{u_{1:k+1}\},\{u_{2:k+2}\},\dots,\{u_{N-k-1:N-1}\}),\label{eq:E21b}
\end{align}
\end{subequations}
where in (\ref{eq:E21b}), we have used the first-order Markov property of $\{U_t\}_{t\in\mathbb{S}}$. Noting the time-homogeneous property of $\{U_t\}_{t\in\mathbb{S}}$ and repeatedly appealing to the probability chain rule in (\ref{eq:E21b}), we get (\ref{eq:E20}). This completes the proof. ~~~~\qed 
\end{pf}
\begin{rem}
\label{R9}
From Theorem \ref{T3}, it is clear that: (i) the sample values of the random variables $\{U_{k+1:N}\}_{k\in\mathbb{N}_0,N\in\mathbb{N}}$ is an ordered sequence constructed from $\{u_{1:N}\}_{N\in\mathbb{N}}$; (ii) the probability distribution of the sequence $\{U_{k+1:N}\}_{k\in\mathbb{N}_0,N\in\mathbb{N}}$ given in (\ref{eq:E20}) is uniquely defined by $P_{\Pi}$ and $P_{\Gamma}$.
\end{rem}
Using Definition \ref{D1} and Theorem \ref{T3}, (\ref{eq:E17}) can be reformulated to the following stochastic programming problem
\begin{subequations}
\label{eq:E23}
\begin{align}
&\tilde{\psi}(U^\star_{k+1:N})=\argmin_{P_{\Pi},P_{\Gamma}}\left\{\sum_{t=1}^{k+1}\Phi(\mathbb{E}_{P_{\Gamma}}[\tilde{L}^\theta_{{t}}(\{U_{k+1})])+\right.\nonumber\\
&~~~~~~~~~~~~~~~~~~~~~~~\left.\sum_{t=k+2}^{N}\Phi(\mathbb{E}_{P_{\Gamma,\Pi}^{k+1:t}}[\tilde{L}^\theta_{{t}}(U_{k+1:t})])\right\}\label{eq:E23a}\\
&\text{s.t. }~~~~0\leq P_{\Pi}(s_i|s_j)\leq1\quad~ \forall~1\leq i,j\leq r^{k+1},\\
&~~~~~~\sum_{i=1}^{r^{k+1}}P_{\Pi}(s_i|s_j)=1 ~~~\quad \forall~1\leq j\leq r^{k+1},\\
&~~~~~~~~~0\leq P_{\Gamma}(s_i)\leq1\quad ~~~~\forall~1\leq i\leq r^{k+1},\\
&~~~~~~\sum_{i=1}^{r^{k+1}}P_{\Gamma}(s_i)=1.
\end{align} 
\end{subequations} 
The expectations in (\ref{eq:E23a}), with respect to ${P}_{\Gamma}$ and ${P}_{\Gamma,\Pi}^{k+1:t}$ can again be approximated using MC sampling, such that
\begin{align}
\label{eq:E26}
\tilde{P}_{\Gamma,\Pi}^{k+1:t}&=\frac{1}{M_u}\sum_{i=1}^{M_u}\delta_{U^i_{k+1:t}}(U_{k+1:t})
\end{align}  
where $\tilde{P}_{\Gamma,\Pi}^{k+1:t}$ is the $M_u$-sample MC estimate. Note that marginalizing (\ref{eq:E26}) with respect to ${\{U_{k+2:N}\}_{k\in\mathbb{N}_0,N\in\mathbb{N}}}$ yields ${\tilde{P}_{\Gamma}=\frac{1}{M_u}\sum_{i=1}^{M_u}\delta_{U^i_{k+1}}(U_{k+1})}$, where $\tilde{P}_{\Gamma}$ is a MC estimate of $P_\Gamma$.  Substituting $\tilde{P}_{\Gamma,\Pi}^{k+1:t}$ and $\tilde{P}_{\Gamma}$ into (\ref{eq:E23a}) yields
\begin{subequations}
\label{eq:E27}
\begin{align}
&\overline{\psi}(U^\star_{k+1:N})=\argmin_{P_{\Pi},P_{\Gamma}}\frac{1}{M_u}\left\{\sum_{t=1}^{k+1}\Phi\left(\sum_{i=1}^{M_u}\tilde{L}^\theta_{{t}}(U^i_{k+1})\right)+\right.\nonumber\\
&~~~~~~~~~~~~~~~~~~~~\left.\sum_{t=k+2}^{N}\Phi\left(\sum_{i=1}^{M_u}\tilde{L}^\theta_{{t}}(U^i_{k+1:t})\right)\right\}\label{eq:E27a}\\
&\text{s.t. }~~~~0\leq P_{\Pi}(s_i|s_j)\leq1\quad~ \forall~1\leq i,j\leq r^{k+1},\\
&~~~~~~\sum_{i=1}^{r^{k+1}}P_{\Pi}(s_i|s_j)=1 ~~~\quad \forall~1\leq j\leq r^{k+1},\\
&~~~~~~~~~0\leq P_{\Gamma}(s_i)\leq1\quad ~~~~\forall~1\leq i\leq r^{k+1},\\
&~~~~~~\sum_{i=1}^{r^{k+1}}P_{\Gamma}(s_i)=1.
\end{align}
\end{subequations}
Note that solving (\ref{eq:E27}), yields ${U^\star_{k+1:N}\sim P_{\Gamma^\star,\Pi^\star}^{k+1:N}}$, which is the optimal distribution of the input sequence.
\begin{cor}
\label{C2}
Let $\overline{\psi}(U^\star_{k+1:N})$ and $\tilde{\psi}(U^\star_{k+1:N})$ be the optimal utility functions, computed by solving the optimization problem in (\ref{eq:E23a}) and (\ref{eq:E27a}), respectively, then 
\begin{equation}
\label{eq:E28}
\overline{\psi}(U^\star_{k+1:N})\xrightarrow[M_u\rightarrow +\infty] {a.s.}\tilde{\psi}(U^\star_{k+1:N}),
\end{equation}
where $\xrightarrow{a.s.}$ denotes almost sure convergence.
\end{cor} 
\begin{pf}
Proof is similar to Theorem \ref{T4}.~~~~~~~~~~~~~~~~~~~~~~~~~~~~~\qed
\end{pf}
\begin{rem}
\label{R10}
There are several advantages of using the formulation given in (\ref{eq:E27}): (a) the optimization is independent of $N\in\mathbb{N}$, as the number of parameters to be estimated are $r^{k+1}(1+r^{k+1})$; (b) easy to include magnitude and other transition constraints on the inputs; and (c) samples from the optimal distribution can be easily sampled, and implemented using a PID or any classical regulator. 
~~~\qed
\end{rem}
In this paper, the optimization problem in (\ref{eq:E27}) is implemented through an iterative approach, that involves standard numerical solvers (\cite{N2006}). The proposed method for input design, including the iterations in the optimization, is summarized in Algorithm \ref{alg:Algo1}.
\begin{algorithm}[!t]
  \caption{Bayesian input design for identification}
  \label{alg:Algo1}
  \begin{algorithmic}[1]
    \STATE Choose an initial value for the input design parameters ${P_{\Gamma}=P^{(0)}_{\Gamma}}$ and ${P_{\Pi}=P^{(0)}_{\Pi}}$. Set ${c \leftarrow 0}$.
    \WHILE{ converged}
    \FOR{$i=1$ to $M_u$}
    \STATE Generate a random input sequence $U^i_{k+1:N}\sim P^{k+1:N}_{\Gamma,\Pi}$ using the distribution given in (\ref{eq:E20}).
    \STATE Generate $M$ random samples of states and parameters from the prior density $\{Z_0^j\}_{j=1}^M\sim p(z_0)$.
    \FOR{$t=1$ to $N$}
    \STATE Generate $M$ random samples of the process states $\{X_t^j|({z^j_{t-1},u^i_{t-1})}\}_{j=1}^M\sim p(x_t|z^j_{t-1},u^i_{t-1})$ and parameters ${\{\theta^j_t=\theta^j_{t-1}\}_{j=1}^M}$ using (\ref{eq:E3}).
    \STATE Generate $M$ random samples of the measurements $\{Y_t^j|({z^j_{t},u^i_{t})}\}_{j=1}^M\sim p(y_t|z^j_{t},u^i_{t})$ using (\ref{eq:E3}).
    \STATE Approximate the lower bound $\tilde{L}_t^\theta$ using (\ref{eq:E25}).
    \ENDFOR
    \ENDFOR
    \STATE Evaluate the approximate cost function in (\ref{eq:E27a}).
    \STATE Use any standard constrained non-linear optimization algorithm to find a new input design parameters ${P_{\Gamma}=P^{(c)}_{\Gamma}}$ and ${P_{\Pi}=P^{(c)}_{\Pi}}$. Set ${c \leftarrow c+1}$. 
    \ENDWHILE
  \end{algorithmic}
\end{algorithm}
\section{Simulation example}
Consider a process described by the following univariate, and non-stationary stochastic SSM (\cite{Tulsyan2013D})

\begin{subequations}
\label{eq:E28}
\begin{align}
X_{t+1}&=aX_t+\frac{X_t}{b+X_t^2}+u_t+V_t,~~V_t\sim\mathcal{N}(0,Q_t),\label{eq:E28a}\\
Y_t&=cX_t+dX_t^2+W_t,~~\quad\qquad W_t\sim\mathcal{N}(0,R_t),\label{eq:E27b}
\end{align}
\end{subequations}
where $\theta\triangleq[a~b~c~d]$ is a vector of model parameters to be estimated, with ${\theta^\star=[0.8~ 0.7~0.6~0.5]}$ being the true parameter vector. The noise covariances are selected as ${Q_t=0.01}$ and ${R_t=0.01}$, for all ${t\in\mathbb{N}}$. For Bayesian identification, ${\{\theta_{t}=\theta_{t-1}\}_{t\in\mathbb{N}}=\theta}$ in (\ref{eq:E27}) is a random process, with ${Z_t=\{X_t,~\theta_t\}}$, such that ${Z_0\sim\mathcal{N}(z_m,z_c)}$, where $z_m=[1~0.7~0.6~0.5~0.4]$, $z_c=\diag(0.01,~0.01,~0.01,~0.01,~0.01)$. Here we assume that ${u_{min}\leq \{u_t\}_{t\in\mathbb{N}}\leq u_{max}}$, where ${u_{min}=-0.8}$ and ${u_{max}=0.8}$. Starting at ${t=0}$, we are interested in choosing an input sequence ${\{u_{1:N}\}_{N\in\mathbb{N}}}$ that would eventually lead to minimization of the MSE of the parameter estimates, computed using an SMC based Bayesian estimator given in \cite{T2013}. Algorithm \ref{alg:Algo1} was implemented with ${N=100}$, ${M=2000}$, and ${M_u=2000}$. 
For input, we consider Example \ref{E2}, with ${g=2}$, such that ${\mathcal{U}=\{u_{min}, u_{max}\}}$. Here ${\{U_t\}_{t\in\mathbb{N}}=\{u_t\}}$ have the following initial and transition probability
\begin{align*}
&\textbf{Case 1:~}P_{\Gamma}=[p_1~~{1-p_1}],\quad P_{\Pi}=\left[
  \begin{array}{cc}
    p_{1} & 1-p_{1} \\
    1-p_{1}& p_{1}  \\
  \end{array}
\right];\\
&\textbf{Case 2:~}P_{\Gamma}=[p_1~~{1-p_1}],\quad P_{\Pi}=\left[
  \begin{array}{cc}
    p_{1} & 1-p_{1} \\
    1-p_{2}& p_{2}  \\
  \end{array}
\right];\\
&\textbf{Case 3:~}P_{\Gamma}=[p_0~~{1-p_0}],\quad P_{\Pi}=\left[
  \begin{array}{cc}
    p_{1} & 1-p_{1} \\
    1-p_{2}& p_{2}  \\
  \end{array}
\right],
\end{align*}
where $p_{i}$, where $i=\{0,1,2\}$ in Cases $1$ through $3$ are the probabilities. For comparison purposes, we also consider a pseudo-random binary signal, which can be represented as
\begin{align*}
&\textbf{Case 4:~}P_{\Gamma}=[0.5~~~~~0.5],\quad P_{\Pi}=\left[
  \begin{array}{cc}
    ~~0.5~~~~&0.5~~\\
    ~~0.5~~~~&0.5~~\\
  \end{array}
\right].
\end{align*}
For all of the above cases, $\Phi(\cdot)$ in (\ref{eq:E27a}) was selected as the trace. Table 1 gives ${P_{\Gamma^\star}}$ and ${P_{\Pi^\star}}$ for Cases 1 through 3 as computed by Algorithm \ref{alg:Algo1}, and Figure 1(a) gives the corresponding trace of the lower bound. It is clear from Table 1 and Figure 1(a) that Case 3 yields the lowest objective function value. Although the objective function value for Case 2 is comparable to Case 3, note that Case 3 provides the most general form of the Markov chain in $\mathcal{U}$. 

Figure 1(b) validates the quality of the designed inputs based on the performance of the Bayesian estimator. From Figure 1(b), it is clear that with Case 3, the estimator yields the lowest trace of MSE at all sampling time. The same is also evident from Table 1; wherein, the sum of the trace of MSE is smallest with Case 3 as the input. The Results are based on $500$ MC simulations, starting with $500$ i.i.d. input path trajectories generated from ${\{U_{1:N}\}\sim P^{1:N}_{\Gamma^\star,\Pi^\star}}$ for Cases 1 through 4. If required, a more rigorous validation of the designed input can be performed using the approach proposed in \cite{Tulsyan2013D}. 

The results appear promising; however, we faced problems in solving the optimization. As discussed earlier, (\ref{eq:E27}) is a stochastic programming problem, as a result (\ref{eq:E27a}) tends to be non-smooth, and have many local minima. In future, we will consider use of stochastic gradient-based methods.
\begin{figure}
\label{figure}
\includegraphics[scale=0.65]{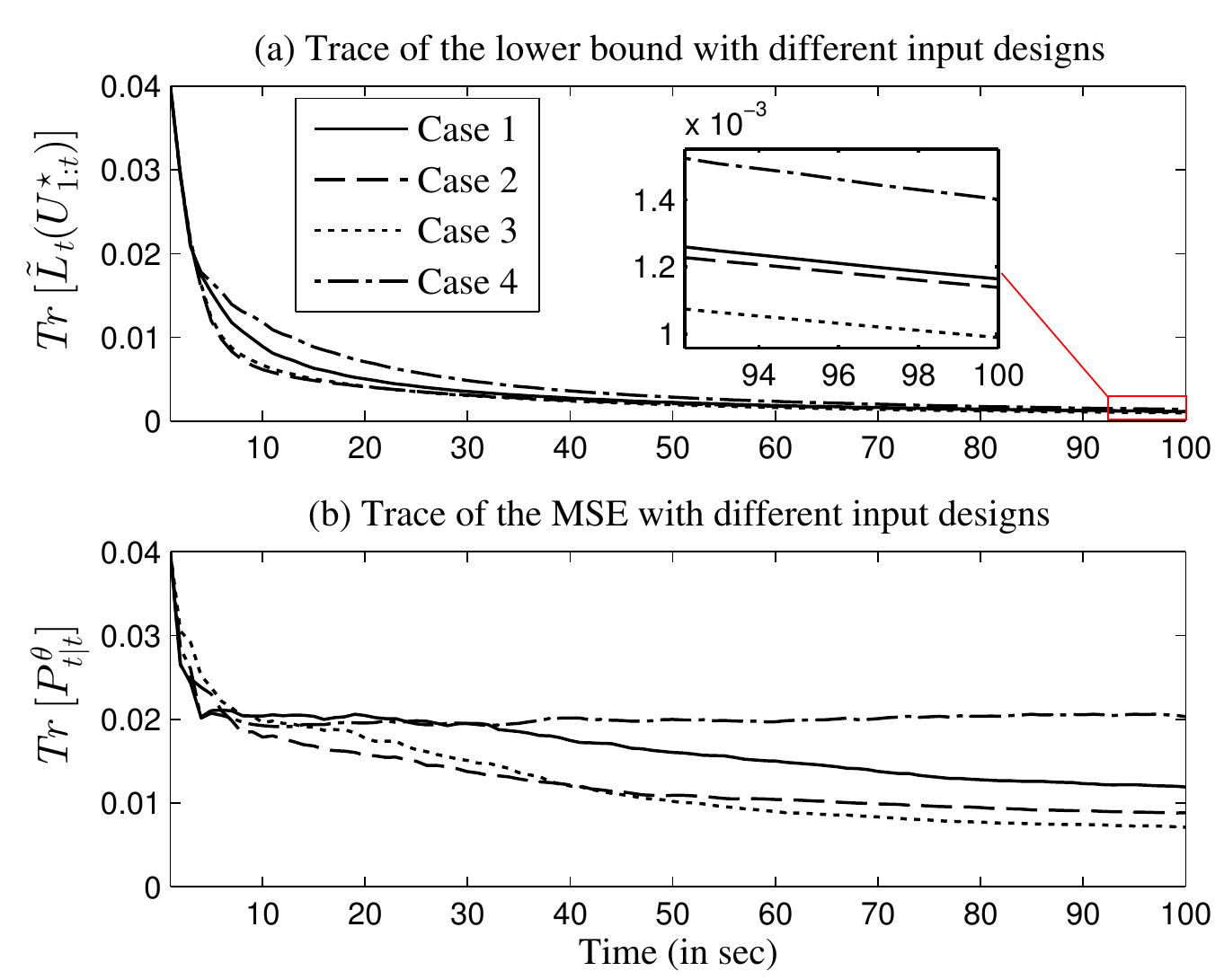} 
\caption{\small{(a) Trace of the approximate lower bound; (b) trace of the MSE. Magnification of the key region of (a) is provided as inset.}}
\end{figure}
\begin{table}[h]
\label{tab:T1}
\caption{\small{Results as computed by Algorithm \ref{alg:Algo1}.}} 
\begin{center}
\begin{tabular}{|c|c|c|c|c|}
\hline 
& Case 1 & Case 2 &  Case 3 & Case 4 \\ 
\hline 
$p_0$&N.A.  & N.A.& $0.34$ & N.A.\\ 
$p_1$& $0.62$& $0.63$& $0.61$& N.A.\\
$p_2$& N.A.&$0.92$ & $0.72$ & N.A.\\
\hline \\[-1em]
$\overline{\psi}(U^\star_{1:100})$ & 0.42 & 0.37 & 0.36 & 0.51 \\
\hline\\[-1em]
$\sum_{t=1}^{100}\tr[P_{t|t}^\theta]$ & 1.66 & 1.27 & 1.25 & 2.02 \\ 
\hline 
\end{tabular} 
\end{center}
\end{table}
\section{conclusions}
An algorithm for input design for Bayesian identification of stochastic non-linear SSM is proposed. The developed algorithm is based on minimization of the PCRLB with respect to inputs. One of the distinct advantages of the proposed method is that the designed input is independent of the Bayesian estimator used for identification. Simulation results suggest that the proposed method can be used to deliver accurate inference on the parameter estimates. 
\bibliographystyle{plain}
\bibliography{ifacconf}

\begin{thebibliography}{14}
\providecommand{\natexlab}[1]{#1}
\providecommand{\url}[1]{\texttt{#1}}
\expandafter\ifx\csname urlstyle\endcsname\relax
  \providecommand{\doi}[1]{doi: #1}\else
  \providecommand{\doi}{doi: \begingroup \urlstyle{rm}\Url}\fi

\bibitem[Bergman(2001)]{NB2001}
N.~Bergman.
\newblock \emph{{Sequential Monte Carlo Methods in Practice}}, chapter
  {Posterior Cra\'mer-Rao Bounds for Sequential Estimation}.
\newblock Springer--Verlag, New York, 2001.

\bibitem[Brighenti et~al.(2009)Brighenti, Wahlberg, and Rojas]{B2009}
C.~Brighenti, B.~Wahlberg, and C.R. Rojas.
\newblock {Input design using Markov chains for system identification}.
\newblock In \emph{Proceedings of the 48th IEEE Conference on Decision and
  Control and the 28th Chinese Control Conference}, pages 1557--1562, Shanghai,
  China, 2009.

\bibitem[Chaloner and Verdinelli(1995)]{C1995}
K.~Chaloner and I.~Verdinelli.
\newblock Bayesian experimental design: A review.
\newblock \emph{Statistical Science}, 10\penalty0 (3):\penalty0 273--304, 1995.

\bibitem[Gilks et~al.(1995)Gilks, Richardson, and Spiegelhalter]{G1995}
W.R. Gilks, S.~Richardson, and D.~Spiegelhalter.
\newblock \emph{Markov Chain Monte Carlo in Practice}.
\newblock Chapman \& Hall, 1995.

\bibitem[Huan and Marzouk(2012)]{H2012}
X.~Huan and Y.M. Marzouk.
\newblock {Simulation-based optimal Bayesian experimental design for non-linear
  systems}.
\newblock \emph{Journal of Computational Physics}, 232\penalty0 (1):\penalty0
  288--317, 2012.

\bibitem[Kantas et~al.(2009)Kantas, Doucet, Singh, and Maciejowski]{Kantas2009}
N.~Kantas, A.~Doucet, S.S. Singh, and J.~Maciejowski.
\newblock {An overview of sequential Monte Carlo methods for parameter
  estimation in general state-space models}.
\newblock In \emph{Proceedings of the 15th IFAC Symposium on System
  Identification}, pages 774--785, Saint-Malo, France, 2009.

\bibitem[K\"uck et~al.(2006)K\"uck, de~Freitas, and Doucet]{K2006}
H.~K\"uck, N.~de~Freitas, and A.~Doucet.
\newblock {SMC samplers for Bayesian optimal non-linear design}.
\newblock In \emph{Proceedings of the Non-linear Statistical Signal Processing
  Workshop}, pages 99--102, Cambridge, U.K., 2006.

\bibitem[Lindley(1956)]{L1956}
D.V. Lindley.
\newblock On a measure of the information provided by an experiment.
\newblock \emph{The Annals of Mathematical Statistics}, 27\penalty0
  (4):\penalty0 986--1005, 1956.

\bibitem[M{\"u}ller and Parmigiani(1995)]{M1995}
P.~M{\"u}ller and G.~Parmigiani.
\newblock {Optimal design via curve fitting of Monte Carlo experiments}.
\newblock \emph{Journal of the American Statistical Association}, 90\penalty0
  (432):\penalty0 1322--1330, 1995.

\bibitem[Nocedal and Wright(2006)]{N2006}
J.~Nocedal and S.J. Wright.
\newblock \emph{Numerical Optimization: Springer Series in Operations
  Research}.
\newblock Springer, New York, 2006.

\bibitem[Tichavsk\'y et~al.(1998)Tichavsk\'y, Muravchik, and Nehorai]{T1998}
P.~Tichavsk\'y, C.~Muravchik, and A.~Nehorai.
\newblock {Posterior Cram\'er-Rao bounds for discrete-time non-linear
  filtering}.
\newblock \emph{IEEE Transactions on Signal Processing}, 46\penalty0
  (5):\penalty0 1386--1396, 1998.

\bibitem[Tulsyan et~al.(2012)Tulsyan, Forbes, and Huang]{T2012a}
A.~Tulsyan, J.F. Forbes, and B.~Huang.
\newblock {Designing priors for robust Bayesian optimal experimental design}.
\newblock \emph{Journal of Process Control}, 22\penalty0 (2):\penalty0
  450--462, 2012.

\bibitem[Tulsyan et~al.(2013{\natexlab{a}})Tulsyan, Huang, Gopaluni, and
  Forbes]{T2013}
A.~Tulsyan, B.~Huang, R.B. Gopaluni, and J.F. Forbes.
\newblock {On simultaneous on-line state and parameter estimation in non-linear
  state-space models}.
\newblock \emph{Journal of Process Control}, 23\penalty0 (4):\penalty0
  516--526, 2013{\natexlab{a}}.

\bibitem[Tulsyan et~al.(2013{\natexlab{b}})Tulsyan, Huang, Gopaluni, and
  Forbes]{Tulsyan2013D}
A.~Tulsyan, B.~Huang, R.B. Gopaluni, and J.F. Forbes.
\newblock {Bayesian identification of non-linear state-space models: Part
  II-Error Analysis}.
\newblock In \emph{Proceedings of the 10th International Symposium on Dynamics
  and Control of Process Systems}, Mumbai, India, 2013{\natexlab{b}}.

\end{thebibliography}
\end{document}